# Physical Layer Security for 6G Systems
## *why it is needed and how to make it happen*

Arsenia Chorti

ETIS UMR 8051 / CY Paris University, ENSEA, CNRS, 95000, Cergy FR

arsenia.chorti@ensea.fr

## I. INTRODUCTION

Sixth generations (6G) systems will be required to meet diverse constraints in an integrated ground-air-space global network. In particular, meeting overly aggressive latency constraints, operating in massive connectivity regimes, with low energy footprint and low computational effort, while providing explicit security guarantees, can be challenging [1]. In addition, the extensive introduction of artificial intelligence (AI) and machine learning (ML) and the rapid advances in quantum computing are further developments that will increase the attack surface of 6G systems [2], [3]. Quite importantly, the massive scale deployment of low-end Internet of things (IoT) nodes, often produced following non-homogeneous production processes, and, importantly, with expected lifespans exceeding 10 years, poses pressing questions with respect to the conceptualization of the overall security architecture.

In this setting, quality of security (QoSec) is envisioned as a flexible security framework for future networks with highly diverse non-functional requirements. Mirroring the differentiated services (DiffServ) networking paradigm, different security levels could be conceptualized, moving away from static security controls, captured currently in zero-trust security architectures [4]. In parallel, the integration of communications and sensing, along with embedded (on-device) AI, can provide the foundations for building autonomous and adaptive security controls [5], orchestrated by a vertical security plane in coordination with a vertical semantic plane [6].

It is in this framework, that we envision the incorporation of physical layer security (PLS) schemes in 6G security protocols, introducing security controls at all layers, for the first time [4]. This exciting prospect does not come, however, without challenges. Despite intense research interest on PLS for more than two decades, it's incorporation in actual security products remains rather elusive, with a few exceptions in terms of RF fingerprinting (physec.de) and multi-factor authentication (silencelaboratories.com). The hurdles to be overcome concern primarily two key issues: the difficulty in providing explicit security guarantees with PLS and the potential degradation in terms of achievable rates. For the incorporation of PLS in 6G, both of these issues need to be addressed. In the following we attempt to draw a generic roadmap in this direction.

## II. PROPOSED ROADMAP FOR INCORPORATING PLS IN 6G

The study of reliable communications involves a typical bottom-up approach; building on idealised source and channel models, powerful theorems on the fundamental limits of achievable rates have been proven. Such results guide the design of actual systems, operating over real transmission channels, given that in practice (a few) connection outages and errors are tolerated. However, the requirements are different in information security, in which the guarantees to be met need to be tight and explicit.

In cryptographic proofs, such explicit guarantees are expressed through the concept of "adversarial advantage", which needs to be negligible (i.e., strictly upper-bounded in polynomial time, reflecting limitations in computational resources) [7]. Related proofs of semantic security (chosen plaintext attack – CPA security, chosen ciphertext attack – CCA security, etc.) are built around games played between legitimate and adversarial parties, when the latter possess specific capabilities, ranging from launching passive attacks (eavesdropping) to launching active attacks (tampering, man-in-the-middle, spoofing). In the cryptographic proofs, a number of idealised mathematical constructions are used, such as pseudorandom number generators (PRNG), pseudorandom functions (PRF) and permutations (PRP) and random oracles (RO), as abstractions of the fundamental cryptographic primitives [7].

Consequently, practical cryptosystems are built to resemble, as closely as possible, the above mentioned idealised mathematical structures. In particular, secure stream ciphers should resemble PRNGs, while block ciphers are built to resemble PRFs or PRPs. An example for the latter, non-linearities are a core element of their design, aiming to induce maximal confusion [8], e.g., with the use of S-boxes. Their design typically involves selected functions from the class of bent functions that are subsequently modified to withstand linear and differential cryptanalysis [9]. As practical crypto primitives are designed to resemble the respective idealised mathematical structures, practical cryptography lingers towards semantic security.

Turning our attention to PLS, information theoretic proofs guaranteeing perfect secrecy or distributed key distillation without leakage, are similarly built using idealised channel models. Importantly, it has been shown that it is possible to prove semantic security for wiretap channels [10], while recently, fundamental results have been published regarding the finite blocklength [11] along with related approaches for analysis and code design [13], [14]. However, it still remains unclear how to translate the theoretical results to practical systems; the key difficulty lies in the fact that the actual transmission channel does not conform to the idealised models used in the proofs. As an example, although the impact of correlations on the secrecy capacity has been largely explored [15], [16] [17], when accounting for more complex phenomena, captured typically through compound or arbitrarily varying channel models, translating channel secrecy capacity expressions [18], [19] to practical code designs becomes tedious. Even more importantly, the adequacy of practical channel models – proposed for example by 3GPP – for information theoretic security is still unclear. We note in passing that for mmWave and sub-THz bands, existing 3GPP models have been shown to have limitations [20], [21]. We therefore need to shed light on how PLS can provide explicit security guarantees in non-idealised transmission conditions.

A first step to resolve related inconsistencies is to learn, e.g., using continuous learning, the channel statistics [22], rather than rely on a set of pre-existing channel models, with the aim to construct secrecy maps [12]. However, the estimation of mutual information from measured CSI vectors, that could enable the practical estimation of the secrecy rate from measured field data, is largely limited in terms of precision by the dataset size [23]. As a result, there are inherent limitations in the online evaluation of secrecy outages directly from field data, e.g., during the network operation.

To bypass this issue, we propose a final step. To engineer the transmission, given the channel model learned, so that the end-to-end transmission resembles transmission over an idealised channel, for which the security guarantees are explicit. A general approach in this direction could exploit reverse Shannon theory [24]; while in the standard Shannon theory a noisy channel is used to simulate a noiseless one, in reverse Shannon theory a noiseless channel is used to simulate a noisy channel, i.e., it is possible to pre-process the transmission so that the end-to-end link resembles a standard wiretap channel. As an example, subsampling (more generally, dimensionality reduction) could serve to simulate a memoryless / block fading channel from a correlated channel. For relevant results for SKG see [25] and [26].

A final point that should be addressed concerns the potential impact of proposed pre-processing approaches on the achievable secrecy or secret key rates. Indeed, in numerous measurement campaigns for SKG [27] or secrecy outage probabilities in subTHz bands [28], it has been shown that in real world scenarios, high entropy (for the extraction of keys) or a consistent SNR advantage (for wiretap coding), cannot be always guaranteed, e.g., for target rates of $0.5$ b/s/Hz. Due to this negative result there has been considerable scepticism with respect to the feasibility of practical PLS systems.

In order to address this issue, we need first identify the target secrecy or secret key rate. In this direction, we propose the employment of PLS for two key goals: (i) for the distillation (using SKG) or distribution (in wiretap channels) of secret keys in hybrid PLS-crypto systems [29], [30]; (ii) for localisation or RF fingerprint based authentication [31], [32]. Consequently, the target rate, is dictated by cryptographic considerations (mode of operation) or by networking parameters (e.g., re-keying due to handovers in high mobility environments). To provide an example for the former, in the record protocol of the transport layer security (TLS) protocol version 1.3, the suggested cipher suite is the `TLS_AES_256_GCM_SHA384 [GCM]`, requiring essentially 96 bytes of key material (32 byte keys, 48 byte MAC secrets, and 2 x 8 byte IVs), while the largest record that can be transmitted amounts to $2^{14}$ bytes [33]. Assuming PLS is used to generate or distribute these keys, then key rates (over multiple subcarriers and antennas) as low as $0.006$ bits/sec could be comfortably targeted, without considering key re-use thanks to zero-round-trip time operation that can actually be extended to incorporate SKG [29]. This toy example showcases that in hybrid PLS-crypto systems the target rates can be extremely small, alleviating concerns related to diminishing throughput.

To summarise, we encapsulate the proposed roadmap for the incorporation of PLS in 6G security protocols in three key ingredients steps:

1) Online, continuous learning of the channel statistics, that will be site specific, context-aware [5], performed by authenticated entities such as trusted bases stations, during the real operation of the network.
2) Transmission and channel engineering of end-to-end PLS links to provide explicit security guarantees, using reverse Shannon theory and tools such as dimensionality reduction.
3) PLS to be used for the distillation and distribution of authentication or symmetric keys in hybrid PLS-crypto systems. Target key rates can be determined by the cryptographic suite of upper layer protocols or network properties (such as handovers).

## REFERENCES

[1] Y. Zou, J. Zhu, X. Wang, and L. Hanzo, "A survey on wireless security: Technical challenges, recent advances, and future trends," *Proc. IEEE*, vol. 104, no. 9, pp. 1727–1765, 2016.

[2] V. Mavroeidis, K. Vishi, M. D. Zych, and A. Jøsang, "The impact of quantum computing on present cryptography," *Int. J. of Adv. Comput. Sci. Appl.*, vol. 9, no. 3, pp. 405–414, 2018.

[3] L. Chen, L. Chen, S. Jordan, Y.-K. Liu, D. Moody, R. Peralta, R. Perlner, and D. Smith-Tone, *Report on post-quantum cryptography*. US Department of Commerce, National Institute of Standards and Technology, 2016, vol. 12.

[4] IEEE INGR Security, "Security and Privacy, International Network Generations Roadmap (INGR) - 2021 Edition."

[5] A. Chorti, A. N. Barreto, S. Kopsel, M. Zoli, M. Chafii, P. Sehier, G. Fettweis, and H. Poor, "Context-aware security for 6G wireless, the role of physical layer security," *IEEE Communications Standards Mag.*, vol. Special Issue on Emerging Security Technologies for 6G Security and Privacy.

[6] A. N. Barreto, S. Ko¨psell, A. Chorti, B. Poettering, J. Jelitto, J. Hesse, J. Boole, K. Rieck, M. Kountouris, D. Singelee, and K. Ashwinee, "Towards intelligent context-aware 6G security," *arXiv:2112.09411*.

[7] D. Boneh and V. Shoup, *A Graduate Course in Applied Cryptography*, Stanford, 2017.

[8] C. E. Shannon, "A mathematical theory of cryptography," September 1, 1945.

[9] K. Nyberg, "Perfect nonlinear s-boxes," in *Advances in Cryptology - EUROCRYPT '91*, Brighton, UK, 1991, p. 378–386.

[10] A. V. M. Bellare, S. Tessaro, "Semantic security for the wiretap channel," in *CRYPTO*. Springer, 2012.

[11] W. Yang, R. F. Schaefer, and H. V. Poor, "Wiretap channels: Nonasymptotic fundamental limits," *IEEE Transactions on Information Theory*, vol. 65, no. 7, pp. 4069–4093, 2019.

[12] Z. Utkovski, M. Frey, P. Agostini, I. Bjelakovic, and S. Stanczak, "Semantic security based secrecy maps for vehicular communications," in *WSA 2021; 25th International ITG Workshop on Smart Antennas*, 2021, pp. 1–5.

[13] M. Shakiba-Herfeh, L. Luzzi, and A. Chorti, "Finite blocklength secrecy analysis of polar and reed-muller codes in bec semi-deterministic wiretap channels," in *2021 IEEE Information Theory Workshop (ITW)*, 2021, pp. 1–6.

[14] M. Shakiba-Herfeh and A. Chorti, "Comparison of short blocklength slepian-wolf coding for key reconciliation," in *2021 IEEE Statistical Signal Processing Workshop (SSP)*, 2021, pp. 111–115.

[15] H. Jeon, N. Kim, J. Choi, H. Lee, and J. Ha, "Bounds on secrecy capacity over correlated ergodic fading channels at high SNR," *IEEE Trans. Inf. Theory*, vol. 57, no. 4, pp. 1975–1983, 2011.

[16] J. Si, Z. Li, J. Cheng, and C. Zhong, "Secrecy performance of multi-antenna wiretap channels with diversity combining over correlated rayleigh fading channels," *IEEE Transactions on Wireless Communications*, vol. 18, no. 1, pp. 444–458, 2019.

[17] M. L. Akter, M. I. Rumman, M. K. Kundu, M. Z. I. Sarkar, and A. S. M. Badrudduza, "Security analysis in correlated nakagami-m fading channels with diversity combining," in *2019 3rd International Conference on Electrical, Computer Telecommunication Engineering (ICECTE)*, 2019, pp. 69–72.

[18] J. No¨tzel, M. Wiese, and H. Boche, "The arbitrarily varying wiretap channel—secret randomness, stability, and super-activation," *IEEE Transactions on Information Theory*, vol. 62, no. 6, pp. 3504–3531, 2016.

[19] A. S. Mansour, H. Boche, and R. F. Schaefer, "List decoding for arbitrarily varying wiretap channels," in *2016 IEEE Conference on Communications and Network Security (CNS)*, 2016, pp. 611–615.

[20] S. Ju and T. S. Rappaport, "Sub-terahertz spatial statistical mimo channel model for urban microcells at 142 ghz," in *2021 IEEE Global Communications Conference (GLOBECOM)*, 2021, pp. 1–6.

[21] Y. Xing, T. S. Rappaport, and A. Ghosh, "Millimeter wave and sub-thz indoor radio propagation channel measurements, models, and comparisons in an office environment," *IEEE Communications Letters*, vol. 25, no. 10, pp. 3151–3155, 2021.

[22] W. Xia, S. Rangan, M. Mezzavilla, A. Lozano, G. Geraci, V. Semkin, and G. Loianno, "Millimeter wave channel modeling via generative neural networks," in *2020 IEEE Globecom Workshops (GC Wkshps*, 2020, pp. 1–6.

[23] D. McAllester and K. Stratos, "Formal limitations on the measurement of mutual information," in *Proceedings of the Twenty Third International Conference on Artificial Intelligence and Statistics*, ser. Proceedings of Machine Learning Research, S. Chiappa and R. Calandra, Eds., vol. 108. PMLR, 26–28 Aug 2020, pp. 875–884.

[24] Z. Luo and I. Devetak, "Channel simulation with quantum side information," *IEEE Transactions on Information Theory*, vol. 55, no. 3, pp. 1331–1342, 2009.

[25] M. Srinivasan, S. Skaperas, and A. Chorti, "On the use of csi for the generation of rf fingerprints and secret keys," *To appear in 25th Int. ITG Workshop on Smart Ant.*, 2021.

[26] M. Srinivasan, S. Skaperas, M. Shakiba-Herfeh, and A. Chorti, "Joint localization based node authentication and secret key generation," in *(to appear) IEEE Int. Conf. Commun. (ICC)*, Seoul, South Korea, May 2022.

[27] P. Walther and T. Strufe, "Blind twins: Siamese networks for non-interactive information reconciliation," in *2020 IEEE 31st Annual International Symposium on Personal, Indoor and Mobile Radio Communications*, 2020, pp. 1–7.

[28] J. Ma, R. Shrestha, J. Adelberg, and *et al.*, "Security and eavesdropping in terahertz wireless links," *Nature*, vol. 563, p. 89–93, 2018.

[29] M. Mitev, A. Chorti, M. Reed, and L. Musavian, "Authenticated secret key generation in delay-constrained wireless systems," *Eurasip J. Wireless Com. Network*, vol. 122, 2020.

[30] ——, "Authenticated secret key generation in delay-constrained wireless systems," *EURASIP J. Wirel. Commun. Netw.*, vol. 2020, pp. 1–29, 2020.

[31] E. Jorswieck, S. Tomasin, and A. Sezgin, "Broadcasting into the uncertainty: Authentication and confidentiality by physical-layer processing," *Proceedings of the IEEE*, vol. 103, no. 10, pp. 1702–1724, 2015.

[32] W. Njima, M. Chafii, A. Chorti, R. M. Shubair, and H. V. Poor, "Indoor localization using data augmentation via selective generative adversarial networks," *IEEE Access*, vol. 9, pp. 98 337–98 347, 2021.

[33] RFC 8446, "The transport layer security (TLS) protocol version 1.3," 2018.